# Phonon-Induced Topological Transition to a Type-II Weyl Semimetal

Lin-Lin Wang[1*], Na Hyun Jo[1,2], Yun Wu[1,2], QuanSheng Wu[3], Adam Kaminski[1,2], Paul C. Canfield[1,2] and Duane D. Johnson[1,2,4]

[1]Ames Laboratory, U.S. Department of Energy, Ames, IA 50011, USA

[2]Department of Physics and Astronomy, Iowa State University, Ames, IA 50011, USA

[3]Theoretical Physics and Station Q Zurich, ETH Zurich, 8093 Zurich, Switzerland

[4]Department of Materials Science and Engineering, Iowa State University, Ames, IA 50011, USA

## Abstract

Given the importance of crystal symmetry for the emergence of topological quantum states, we have studied, as exemplified in NbNiTe$_2$, the interplay of crystal symmetry, atomic displacements (lattice vibration), band degeneracy, and band topology. For NbNiTe$_2$ structure in space group 53 (*Pmna*) – having an inversion center arising from two glide planes and one mirror plane with a 2-fold rotation and screw axis – a full gap opening exists between two band manifolds near the Fermi energy. Upon atomic displacements by optical phonons, the symmetry lowers to space group 28 (*Pma2*), eliminating one glide plane along *c*, the associated rotation and screw axis, and the inversion center. As a result, twenty Weyl points emerge, including four type-II Weyl points in the Γ–X direction at the boundary between a pair of adjacent electron and hole bands. Thus, optical phonons may offer control of the transition to a Weyl fermion state.

# Introduction

For novel topological quantum states to emerge, a certain combination of crystalline symmetry coupled with or without time-reversal symmetry is required. For example, mirror planes can protect surface states in topological crystalline insulators[1-3] (TCIs). More recently, non-symmorphic symmetries, such as glide planes and screw rotations, have been found to give hourglass fermions[4] on insulator surfaces and topological nodal loops[5] in metals. Beyond just searching for crystal structures with these symmetry operations in space groups, one can also inspect the possible symmetry changes that can be induced by atomic basis displacements. As the irreducible representations of the crystal symmetry group, lattice vibrations or optical phonon modes in certain direction can lower the crystal symmetry by breaking a subset of the symmetry operations. A phonon-induced topological transition has been proposed for band inversion driven by electron-phonon coupling that renormalizes band gap.[6-9] Thus far, the effect of phonon-induced symmetry changes on topological transitions has been only studied for the loss of mirror symmetry[10] in TCIs.

Weyl fermions[11] can emerge in systems where either time-reversal or inversion symmetry is broken. The Weyl points (WPs) appear as pairs of opposite chirality or topological charges in momentum space with linear dispersion, acting as monopoles (sinks or sources) for Berry curvature. The projection of WPs on a surface without overlap gives open Fermi arcs. Weyl fermions give rise to anomalous electronic properties, such as unsaturated magnetoresistance. Weyl semimetals (WSM) without inversion symmetry were first realized in TaAs structures,[12-16,17] where spin-orbit coupling (SOC) gaps out the nodal rings in the band structure except at the WPs, creating point-like Fermi surfaces. The discovery of type-II WSM in $WTe_2$ crystal,[18] with WPs emerging at boundaries between electron and hole pockets near the Fermi energy ($E_F$), extended the classification of topological states, because Lorentz invariance necessary in quantum field theory is not required in the solid-state band-structure. More type-II WSMs continue to be found, e.g., $MoTe_2$[19,20], $WP_2$, $MoP_2$[21], $TaIrTe_4$[22], $NbIrTe_4$[23], and $Ta_3S_2$[24].

Although glide, mirror planes, and screw rotations tend to support novel topological states, their combination often yields inversion symmetry. Here we show a system, $NbNiTe_2$ compound with crystal structure in space group 53 (*Pmna*), where Weyl fermions are prohibited due to an inversion center arising from two glide planes and one mirror plane with 2-fold rotation or screw axis. The band structure has a full gap between two band manifolds around $E_F$ and the band topology is trivial. However, after accounting for atomic displacements due to favorable optical phonon modes (mostly Nb-Te stretching), the glide plane along *c* and the associated rotation and screw axis are lost, which breaks inversion symmetry and lowers the symmetry to space group 28 (*Pma2*). As the result, 20 WPs emerge in the whole Brillouin zone (BZ) at the boundary of highest filled band N and lowest empty band N+1 (N=56 is the number of valence electrons in the system). This intriguing interplay of crystalline symmetry, lattice vibrations, band degeneracy and band topology provides an opportunity to induce topological phase transitions by controlling optical phonons.

## Computational Methods

All density functional theory[25,26] (DFT) calculations with(out) SOC were performed with the PBE[27] exchange-correlation functional using a plane-wave basis set and projector augmented wave[28] method, as implemented in the Vienna Atomic Simulation Package[29,30] (VASP). The tight-binding model based on maximally localized Wannier functions[31-33] was constructed to reproduce closely the band structure including SOC in the range of $E_F \pm 1eV$. Then surface Fermi arcs and spectral functions were calculated with the surface Green's function methods[34-38]. Berry curvatures near WPs are calculated based on Wannier interpolation[39]. In the DFT calculations, we used a kinetic energy cut-off of 269.5 eV, Γ-centered Monkhorst-Pack[40] (6×7×6) *k*-point mesh, and Gaussian smearing of 0.05 eV. For structural relaxation, the atomic basis positions are relaxed until the absolute value of force on each atom is less than 0.005 eV/Å with the fixed cell dimensions of *a*=7.955, *b*=6.258 and *c*=7.203 Å from experiment[41,42]. Electronic band dispersions were checked with higher kinetic energy cut-off (500 eV) and with exchange-correlation functionals that includes van der Waals interactions; these changes only shift slightly the location of WPs, leaving our findings unchanged, as expected for a symmetry-induced topologically phenomenon.

## Results and Discussion

For NbNiTe$_2$ and the related compounds, two orthorhombic crystal structures of space group 53 (*Pmna*) and 28 (*Pma2*) have been reported.[41-43] Figure 1(a) and (b) shows the crystal structure of NbNiTe$_2$ in two different views. After swapping the crystallographic *b* and *c* axis for space group 53, i.e., choosing the same *c* axis along the direction of stacked van der Waals layers as in 28, the two structures are almost the same, except for some small atomic basis displacements that result in non-equivalent sites and lowering the space-group symmetry. Along *b* (or *y*) direction, Fig. 1(a), the structure can be viewed as an hcp structure of Te with half of the octahedral sites occupied by Nb and a quarter of the tetrahedral sites by Ni. Along *c* (or *z*) direction in Fig. 1(b), i.e., the direction of stacked van der Waals layers, the sandwiched layer has a zigzag Te double layer on the outside and a layer of Nb atoms and tilted Ni dimer in the inside. The distance of 2.50 Å between Ni atoms in the dimer is close to the nearest-neighbor distance of 2.49 Å in fcc Ni. The only difference between structure 53 and 28 are the small atomic basis displacements, mostly along *c* direction (see the less than ideal positions in Fig. 1(b)). This breaks the glide plane along *c* direction together with the 2-fold rotation along *a* and the 2-fold screw axis along *b* (see Table 1 for the symmetry operations in the two structures). As the result, the inversion center is removed when the symmetry is lowered from space group 53 to 28. Upon relaxation in DFT with the fixed cell dimensions of 28 from experiments[41,42], the different interlayer distances from the small atomic basis displacements disappear, the structure of space group 28 becomes that of 53.

To assess whether such change in atomic basis displacements is accessible via lattice vibration or phonon near room temperature, we list those displacements for the change from space group 53 to 28 in Appendix A. The displacements range from 0.023 to 0.072 Å. The associated restoring forces acting on the atoms after such displacement are shown in Fig. 1(c) with the view in *a* direction. It is not a single optical phonon mode, but a combination of different modes. The largest displacements are from Nb and Te, with the bond length between them decreased from 2.795 to 2.708 Å on one side of the sandwiched layers. This corresponds to the asymmetric optical phonon mode (Nb-Te bond stretching along *yz* direction). From the restoring force and displacement, we estimate the associated phonon mode energies, which are all below 30 meV. Furthermore, the phonon band structure in Fig.1 (d) indicates that all the phonon modes, including the optical ones, are below 35 meV. So the atomic basis displacements required for the topological phase transition in NbNiTe$_2$ (see below) are accessible via optical phonons near room temperature (~26 meV). The soft phonon mode along Γ–Z with negative frequencies corresponds to a small monoclinic distortion of the orthorhombic cell with 3 meV lower in energy. But including zero point energy and at finite temperature, the orthorhombic structure is still more stable.

The effects of lowering symmetry on electronic band structure without and with SOC can be seen in Fig. 2. For structure 53 without SOC (Fig. 2(a)), the bands are at least doubly degenerated due to crystal symmetry beside the spin degeneracy. Between the two band manifolds around the $E_F$ (one is from −0.5 to 0.0 eV and the other 0.0 to 0.5 eV), there are gaps in the BZ except around Γ–X, Γ–Y and Γ–Z. Specifically, along Γ–X from −0.2 to 0.1 eV, there are two hole bands crossing with an electron band. Upon atomic basis displacements along $c$ to reach structure 28 (Fig. 2(b)), the two hole bands shift to slightly lower energy and the electron band moves up in energy by 0.2 eV near Γ, lifting the band crossing with the lower hole band. The segment of the electron band near Γ lies on top of the higher hole band and they are still doubly degenerated beside the spin degeneracy. In contrast, the degeneracy along S-Y and R-T are lifted due to lowered crystal symmetry from 53 to 28. For structure 53 with SOC (Fig. 2(c)), the SOC lift the double degeneracy for most high-symmetry directions, except for Γ–X and a few others. More importantly, with SOC there is a small gap between the two band manifolds throughout the BZ. Along Γ–X from −0.2 to 0.1 eV, the two crossings between the two hole and one electron bands are gapped out and the segment of the electron band near Γ is pushed to lower energy at −0.4 eV. The result is the formation of one hole and one electron band with band inversion closely lying along Γ–X from −0.2 to 0.1 eV (Fig. 2(e)). Now adding atomic basis displacements along $c$ to 53 with SOC (or turning on SOC to 28), the degeneracy due to crystal symmetry and spin in the band structure is lifted except along Γ–Z (Fig.2 (d)). For the electron and hole bands along Γ–X around $E_F$−128 meV (Fig. 2(e)), the lifting of degeneracy gives four bands. Two of the bands are almost touching with orbital characters switching (Fig. 2(f)). This hints the existence of type-II WPs at locations slightly away from Γ–X. The change in the projection on $d_{yz}$ orbital also reflects that the atomic basis displacement is mostly the stretching mode for Nb-Te along $yz$ direction (Fig. 1(c)). The above dispersions with the unique band inversion features persist for other exchange-correlation functionals including van der Waals interactions[44-46] and high kinetic energy cut-offs (see Appendix B).

After constructed a tight-binding model based on maximally localized Wannier functions[31-33] to reproduce closely the band structure of space group 28 including SOC (Fig.2 (d)) in the range of $E_F$±1eV, we have done a thorough search for WPs in the whole BZ and found a total of 20 WPs in 4 different sets; the momentum and energy locations are listed in Table 2 and plotted in Fig.3 (a). The WPs within each set are related by symmetry and translation. These WPs are confirmed by calculating the chirality with Berry curvature. The WPs in set 1 are the ones near the Γ–X at $E_F$−128 meV. The lifting of degeneracy along S-Y and R-T also gives interesting hourglass[4] type of band crossing features at lower and higher energy range. But these band manifolds with hourglass features have gaps between each other, so they are topologically trivial. The hourglass feature within the same band manifold can also host non-symmorphic nodal loops. But the number of valence electrons of 56 for $NbNiTe_2$ does not meet the 4n+2 criteria[5] necessary for a half-filled band manifold near $E_F$ with hourglass features. Thus, there is no non-symmorphic nodal loop in this system either.

From now on, we will focus on the WPs set 1 near the Γ–X. To get a full view of the band structure features near $E_F$, we plot the 3-dimensional bulk Fermi surface in Fig. 3(b). At $E_F$ there is a small hole pocket around Γ. Going along Γ–X, there is an electron pocket centered at X and extended to the midpoint of Γ–X agreeing with the band structure in Fig. 2(d). There are also two connecting hole cylinders along $k_z$ sitting away from Γ–Y. Going to $E_F$−128 meV in Fig.3(c), the hole pocket and cylinders expand significantly and become all connected around Γ. In contrast, the electron pocket shrinks from the side of X point. Effectively the boundaries of electron and hole pockets become closer and start to touch along Γ–X.

To confirm these touching points are indeed the WPs, first we zoom in one of the point just away from Γ–X and plot the contour of the gap between the filled N and empty N+1 band at $k_z$=0 plane in Fig. 3(d). The gap approaching to zero is clearly seen. The linear dispersion around the WP is then plotted in Fig. 3(e). Next to identify the topological index, we plot the Berry curvatures of two neighboring WPs across Γ–X in Fig. 3(f). The sink and source of Berry curvatures confirms they are indeed WPs with Chern number, C=−1 and 1, respectively. Following the same procedure, we also confirmed the WPs in the other three sets (their linear dispersions are shown in Appendix C).

The WPs in set 1 will manifest themselves as Fermi arcs on the (001) surfaces because their projection along (001) are not overlapped. We plot the surface Fermi surface in Fig. 4(a) and (b) based on the surface Green's function method. Due to the atomic basis displacements along c that breaks the glide plane symmetry, the top and bottom surface states are not equivalent anymore. One can clearly see the triangular shape of the electron pockets touching with the large

X-shaped hole pockets that gives the WPs (labeled as red and blue circles) along Γ–X. Zooming in, we show the Fermi arcs connecting the neighboring WPs across Γ–X on the top and bottom (001) surfaces in Fig.4(c) and (d), respectively. The Fermi arc on the top surface is concave with respect to Γ, while that on the bottom surface is convex to Γ.

We also plot the surface spectral functions in Fig. 4(e) and (f) for the top and bottom (001) surfaces along $k_x$ with $k_y=0.0128\times(2\pi/b)$. For the two surface states around $E_F$−120 eV at the top surface (Fig. 4(e)), the lower surface state is originated from the WP at $E_F$−128 meV as indicated by the plus sign. The one surface state from the WP at $E_F$−128 meV at the bottom surface (Fig. 4(f)) is also clearly seen, which connects the electron and hole bands via the WP.

## Conclusion

In summary, without restricting ourselves to searches of non-symmorphic symmetry operations in crystal structures with space groups, we have studied, using NbNiTe$_2$ compound as an example, the interplay of crystalline symmetry, atomic displacements (lattice vibration), band degeneracy, and band topology. For NbNiTe$_2$ in the crystal structure of space group 53 (*Pmna*), the band structure (with spin-orbit coupling) shows a full gap opening between two band manifolds near the Fermi energy. Upon atomic displacements, accessible via optical phonons around room temperature, the symmetry can be lowered to space group 28 (*Pma2*). A total of 20 Weyl points emerge, including four type-II Weyl points in the Γ–X direction at the boundary between a pair of touching electron and hole pockets. Our study opens up the opportunity to induce topological phase transitions to Weyl fermions by optical phonons. Such a phonon-induced transition will give rise to interesting electron-phonon coupling effects on transport, detectable in temperature-dependent dynamic transport measurement by ultrafast laser spectroscopy, which is in our future research plans.

## Acknowledgement

This work was supported by the U.S. Department of Energy (DOE), Office of Science, Basic Energy Sciences, Materials Science and Engineering Division, and, for LLW, by Ames Lab's laboratory-directed research and development (LDRD) program. Work was performed at Ames Laboratory, which is operated by Iowa State University under contract DE-AC02-07CH11358. NHJ was supported by the Gordon and Betty Moore Foundation's EPiQS Initiative through Grant GBMF4411. YW was supported by CEM, a NSF MRSEC, under Grant No. DMR-1420451. QSW was funded from Microsoft Research and the Swiss National Science Foundation through the National Competence Centers in Research MARVEL and QSIT.

* llw@ameslab.gov

| Symmetry operation | 53 | 28 |
|---|---|---|
| Along *c* | 2-fold rotation, glide plane with ½ *b* shift | 2-fold rotation |
| Along *a* | 2-fold rotation, mirror plane | mirror plane |
| Along *b* | 2-fold screw axis, glide plane with ½ *a* shift | glide plane with ½ *a* shift |
| Inversion center | Yes | No |

Table 1. Symmetry operations for orthorhombic NbNiTe$_2$ with space group 53 (*Pmna*) and 28 (*Pma2*).

| Set | i | $k_x, k_y, k_z$ ($2\pi/a, 2\pi/b, 2\pi/c$) | E (eV) | C |
|---|---|---|---|---|
| 1 | 1 | +0.3160, +0.0128, 0 | −0.128 | −1 |
|   | 2 | +0.3160, −0.0128, 0 |        | +1 |
|   | 3 | −0.3160, +0.0128, 0 |        | +1 |
|   | 4 | −0.3160, −0.0128, 0 |        | −1 |
| 2 | 1 | +0.3103, +0.2705, 0.5 | +0.055 | −1 |
|   | 2 | +0.3103, −0.2705, 0.5 |        | +1 |
|   | 3 | −0.3103, +0.2705, 0.5 |        | +1 |
|   | 4 | −0.3103, −0.2705, 0.5 |        | −1 |
| 3 | 1 | +0.3523, +0.2622, 0.5 | +0.046 | +1 |
|   | 2 | +0.3523, −0.2622, 0.5 |        | −1 |
|   | 3 | −0.3523, +0.2622, 0.5 |        | −1 |
|   | 4 | −0.3523, −0.2622, 0.5 |        | +1 |
| 4 | 1 | +0.4447, +0.1764, +0.1522 | −0.086 | +1 |
|   | 2 | +0.4447, −0.1764, +0.1522 |        | −1 |
|   | 3 | −0.4447, +0.1764, +0.1522 |        | −1 |
|   | 4 | −0.4447, −0.1764, +0.1522 |        | +1 |
|   | 5 | +0.4447, +0.1764, −0.1522 |        | +1 |
|   | 6 | +0.4447, −0.1764, −0.1522 |        | −1 |
|   | 7 | −0.4447, +0.1764, −0.1522 |        | −1 |
|   | 8 | −0.4447, −0.1764, −0.1522 |        | +1 |

Table 2. Momentum ($k_x, k_y, k_z$), energy (E) locations and the chirality (C), of all 20 WPs in four different sets in the whole BZ for NbNiTe$_2$ in space group 28 (*Pma2*), as also shown in Fig.3(a).

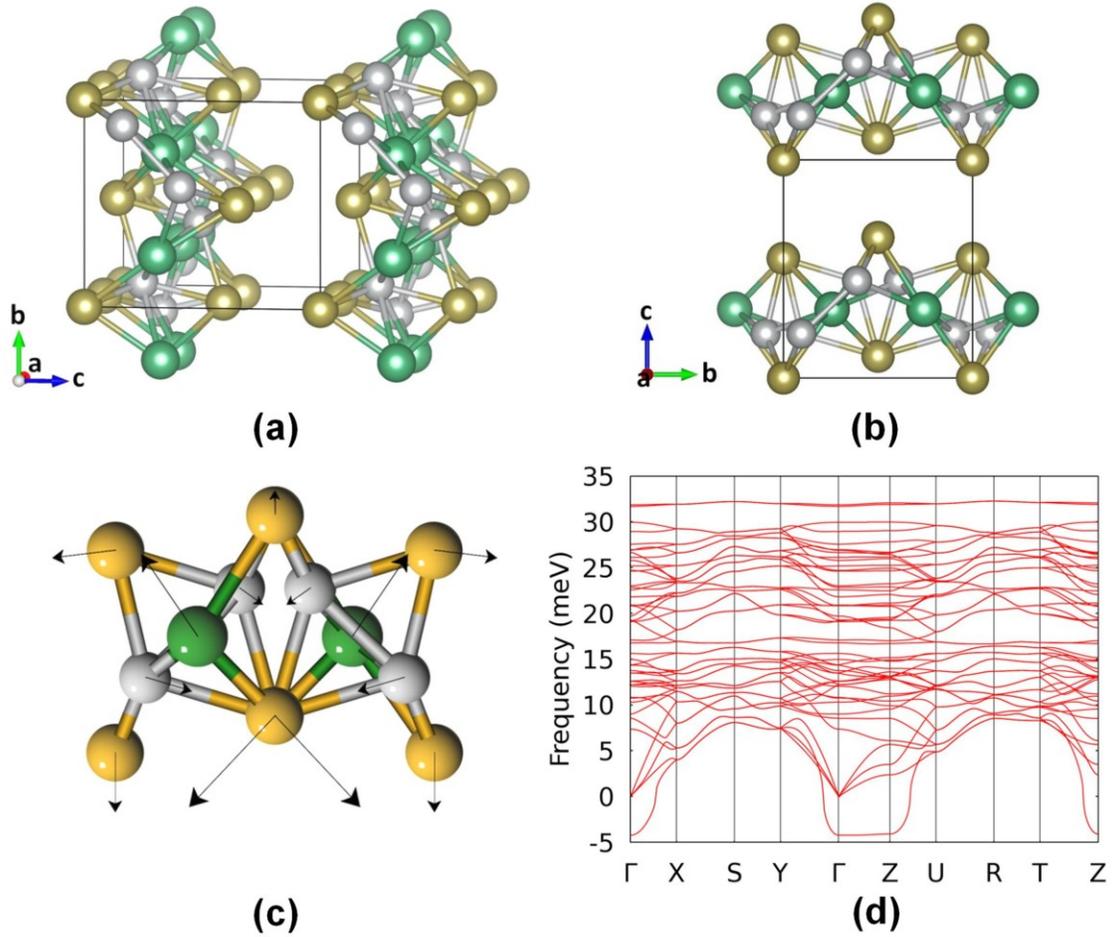

Figure 1. (a) Orthorhombic structures of NbNiTe$_2$ in space group 53 (*Pmna*) and 28 (*Pma2*), where Table 1 lists their different symmetry operations. Green, gray and brown spheres are for Nb, Ni and Te, respectively. (b) Structure viewed along *a* (*yz* plane). (c) Atomic structure viewed along *a* showing restoring forces (vectors) for the change from space group 53 to 28. The magnitude of displacements and estimate of phonon energies are listed in Table S1. (d) Phonon band structure for *Pmna*, revealing an unstable mode with negative frequencies.

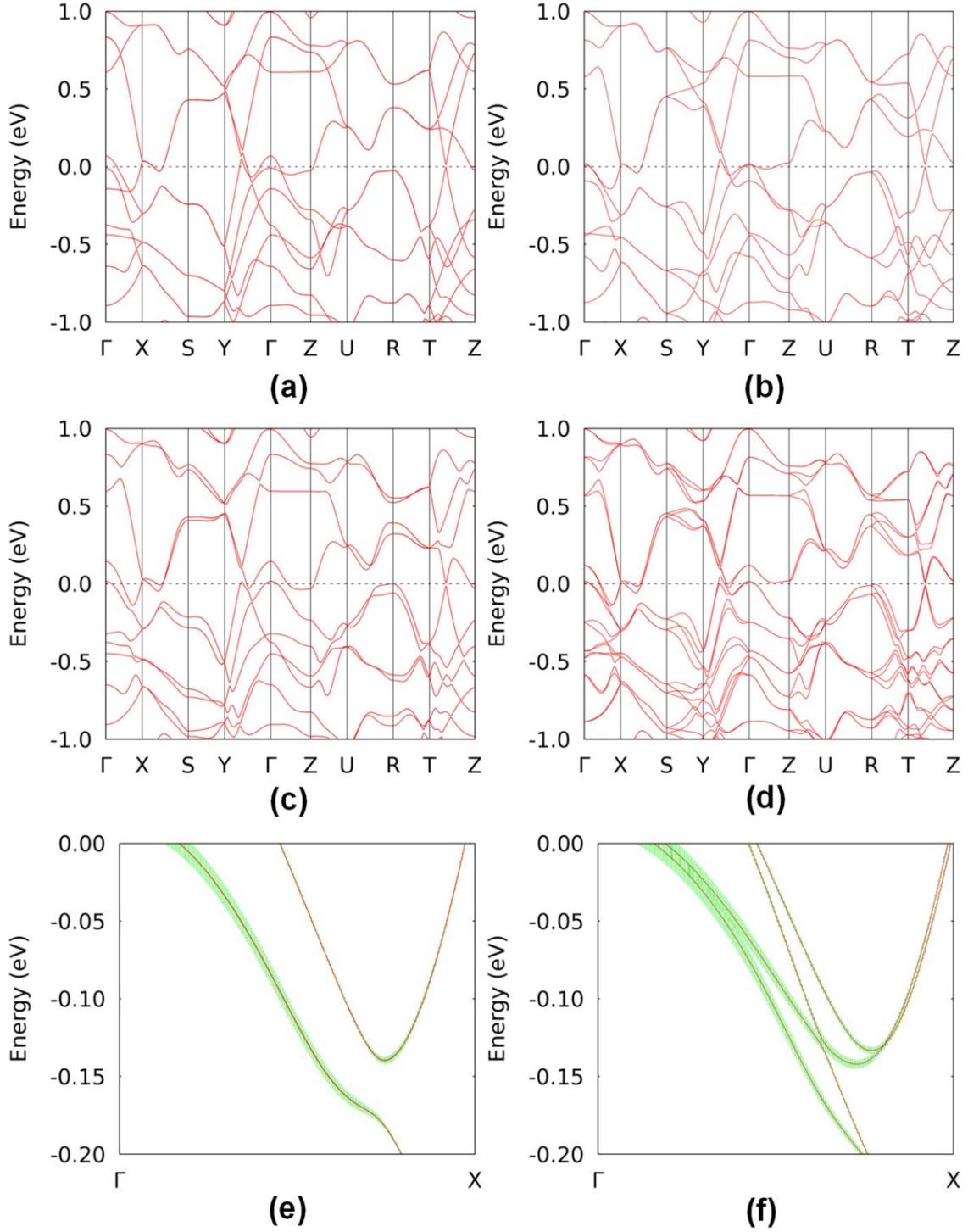

Figure 2. Band structures of space group 53 and 28 without SOC [(a) and (b), respectively], and with SOC [(c) and (d), respectively]. (e) and (f) Band structures zoomed in from (c) and (d), respectively, along Γ–X near $E_F$, with projection on $d_{yz}$ orbital shown by green dots.

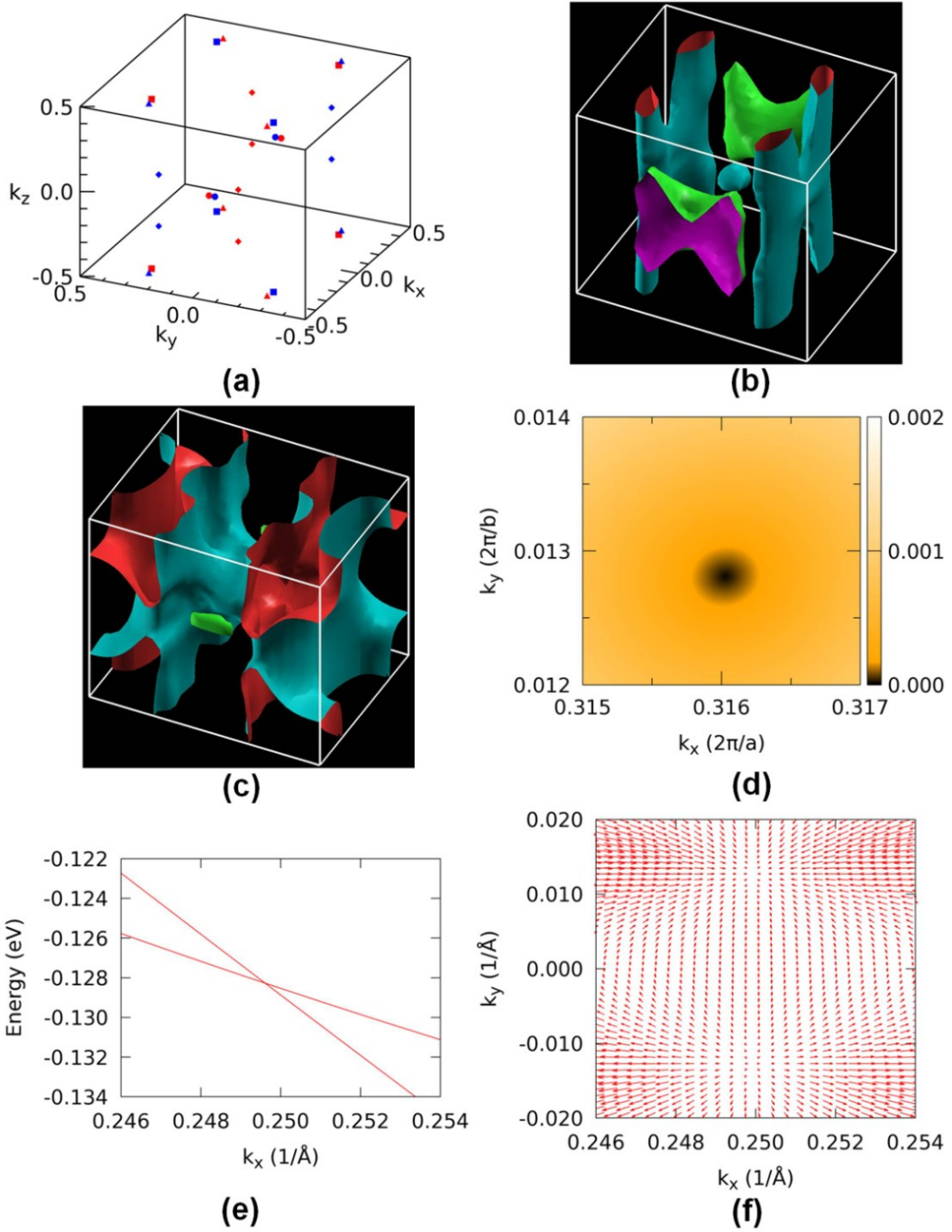

Figure 3. (a) Four sets of WPs in the whole BZ for NbNiTe$_2$ in space group 28 (*Pma2*) as listed in Table 2: circles are set 1, squares set 2, triangles set 3 and diamonds set 4. The blue (red) color is for Chern number C=−1 (1). (b) Fermi surfaces at E$_F$ and (c) Fermi surfaces at E$_F$−100 meV. Aquamarine (green) is for hole (electron) bands. (d) Contour of the gap between band N (56) and N+1 (57) near the WP in set 1 (0.3160, 0.0128, 0) (e) Band dispersion along $k_x$ near the same WP (0.2496, 0.0129, 0) in the unit of 1/Å. (f) Berry curvature near the pair of WP in set 1 across Γ–X. Chern number C=−1 (1) for sink (source) of Berry curvature.

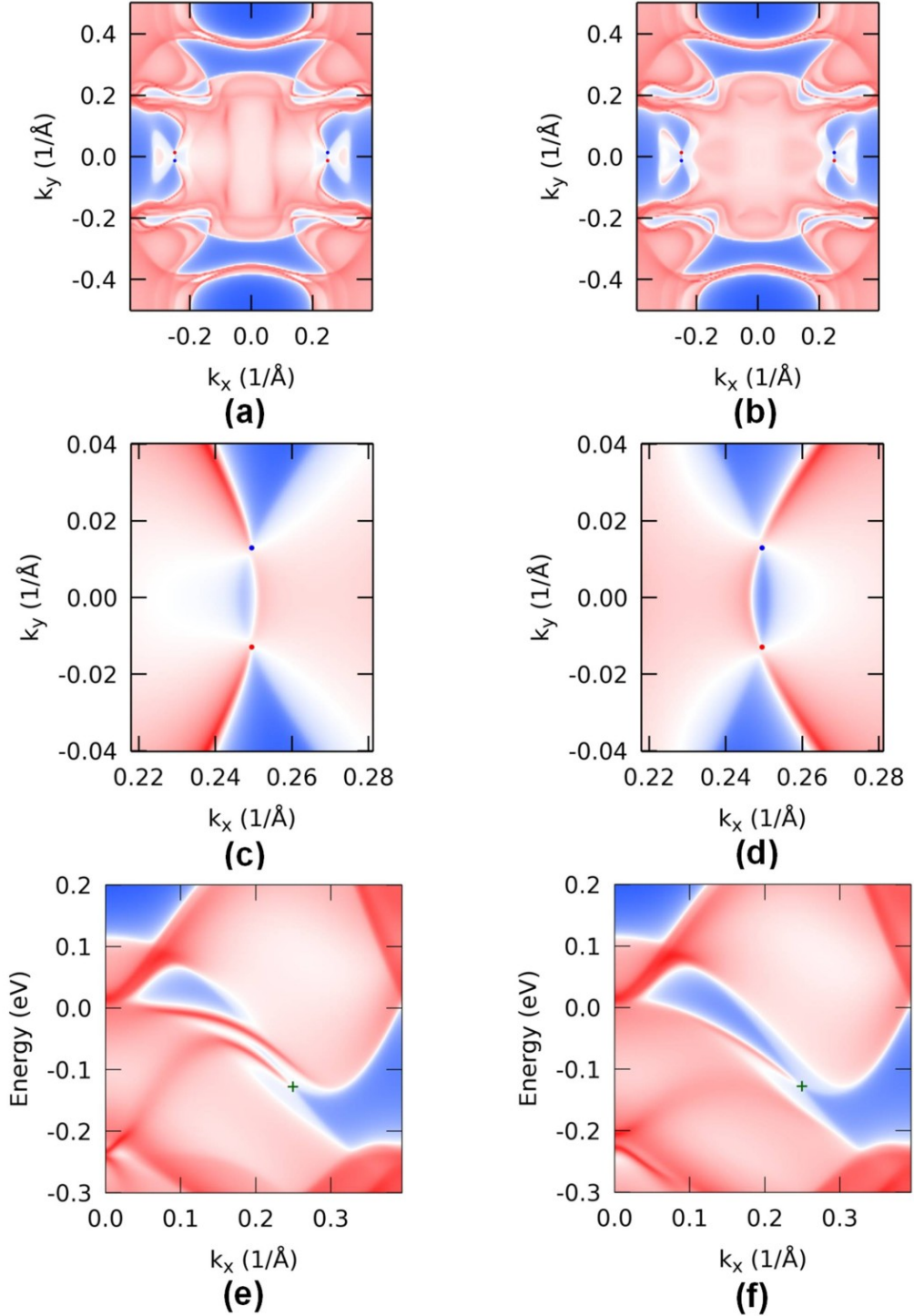

Figure 4. Fermi arcs at $E_F-128$ meV on (a) top and (b) bottom (001) surfaces and the features zoomed in near WP projections in (c) and (d); and surface spectral functions along $k_x$ with $k_y=0.0128\times(2\pi/b)$ on (e) top and (f) bottom (001) surfaces. Low, medium, and high density of states is indicated by blue, white and red colors, respectively. The locations of the WP projections are labeled by red and blue circles in (a)-(d), and green plus signs in (e) and (f).

# Appendix A: Vibrational Energy Estimation

To assess whether the change in atomic basis displacements from space group 53 to 28 is accessible via lattice vibration or phonon near room temperature, we list those in Table S1. The displacements range from 0.023 to 0.072 Å. The associated restoring forces acting on the atoms after such displacement are shown in Fig. 1(c) with the view in *a* direction.

| Site | Displacement (Å) | Force (eV/Å) | Force constant (eV/Å$^2$) | Vib. Energy (eV) |
|---|---|---|---|---|
| Nb (4d) | 0.067 | 0.899 | 13.49 | 0.030 |
| Ni (2c) | 0.072 | 0.377 | 5.23 | 0.014 |
| Ni (2c) | 0.038 | 0.188 | 5.01 | 0.004 |
| Te (2a) | 0.033 | 0.490 | 15.05 | 0.008 |
| Te (2c) | 0.044 | 1.130 | 25.91 | 0.025 |
| Te (2c) | 0.013 | 0.530 | 41.89 | 0.003 |
| Te (2b) | 0.023 | 0.141 | 6.09 | 0.002 |

Table 3. For the structural change NbNiTe$_2$ from space group 53 to 28 via lattice vibration, listed are the non-equivalent atomic sites in space group 28, displacements, forces, and the estimation of force constants and the corresponding vibrational or phonon mode energies.

# Appendix B: Different Exchange-Correlation Functionals

The robustness of the topological features in the band structures with SOC for NbNiTe$_2$ in space group 28 has been checked with different exchange-correlation functionals in DFT.

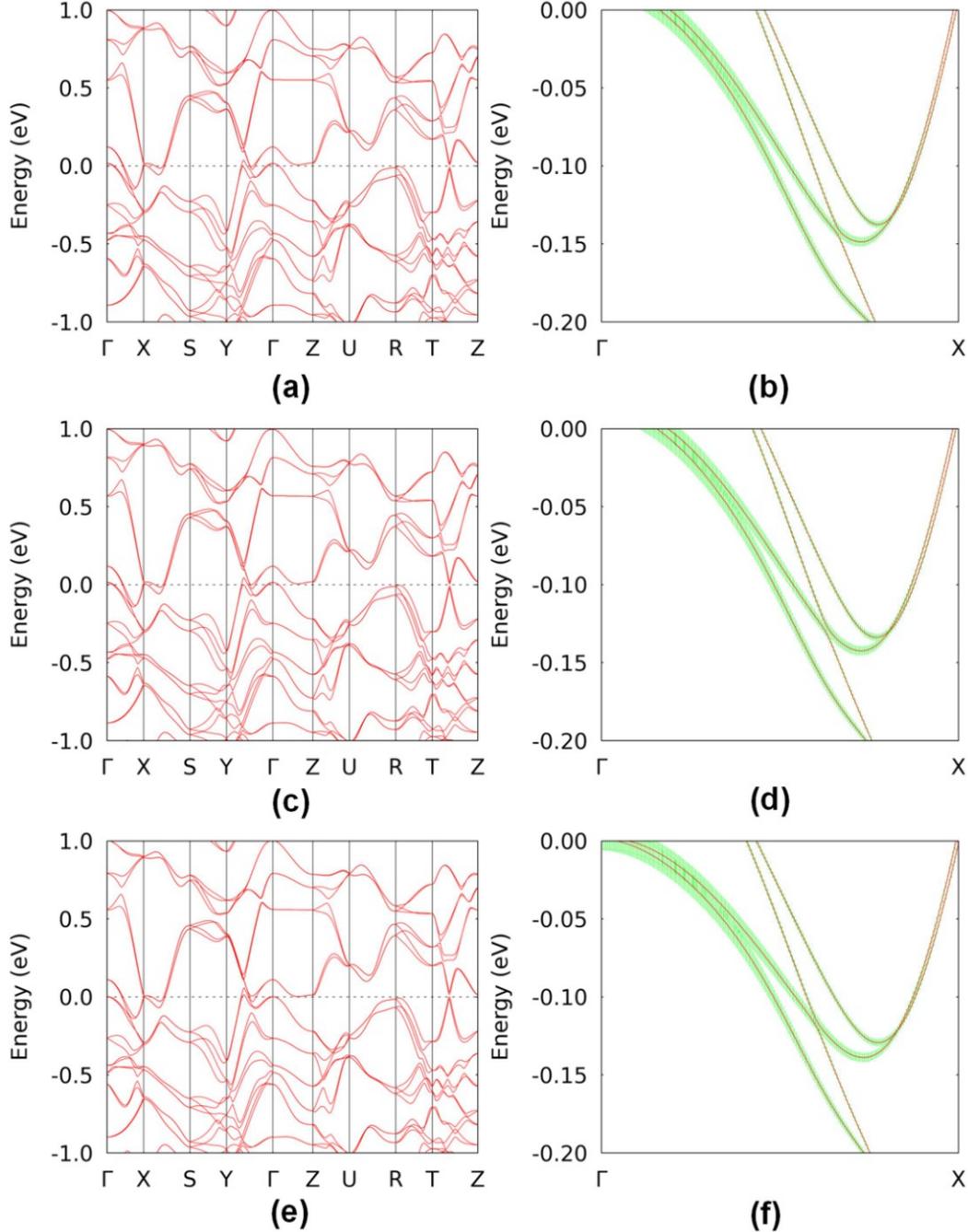

Figure 5. Band structures with SOC for NbNiTe$_2$ in space group 28 calculated in LDA ((a) and (b)), with van der Waals interactions of empirical DFT-D2 (Ref.44) ((c) and (d)), and density functional DF1-optPBE (Refs.45,46) ((e) and (f)). Green dots are the projection on $d_{yz}$ orbital.

# Appendix C: Linear Band Dispersion at Other WPs

The linear band dispersion at the WPs in Set 2, 3 and 4 for NbNiTe$_2$ in space group 28 are presented below for completeness.

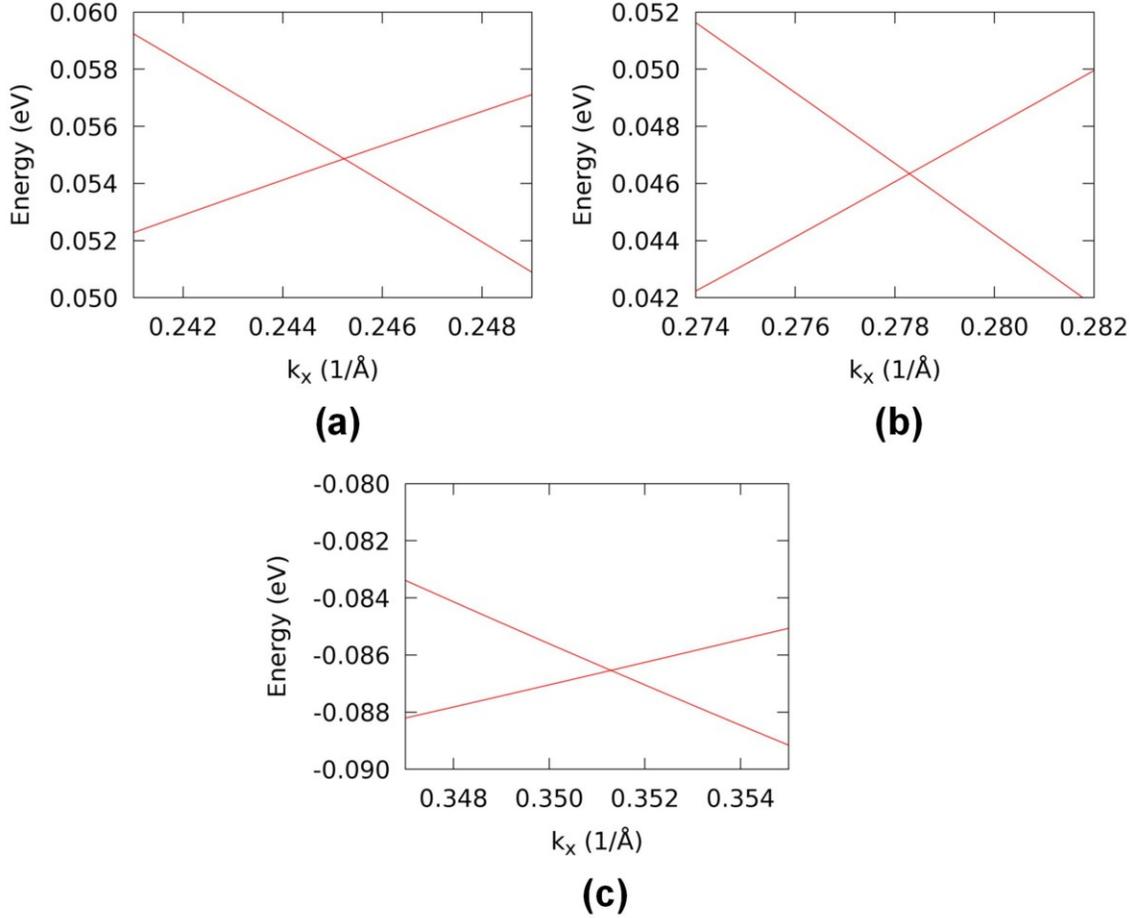

Figure 6. Band dispersion along $k_x$ near the WPs for (a) Set 2 (0.3103, 0.2705, 0.5) or (0.2451, 0.2716, 0.4359) in the unit of 1/Å, (b) Set 3 (0.3523, 0.2622, 0.5) or (0.2783, 0.2633, 0.4361) in the unit of 1/Å, and (c) Set 3 (0.4447, 0.1764, 0.1522) or (0.3513, 0.1771, 0.1328) in the unit of 1/Å.